\documentclass[aps,prl,twocolumn]{revtex4}
\usepackage[usenames]{color}
\newcommand{\comment}[1]{}
\usepackage{graphicx}
\usepackage{epsfig}

\newcommand{\ket}[1]{| {#1} \rangle}
\newcommand{\expect}[1]{\langle {#1} \rangle}
 
\begin{document}

\bibliographystyle{revtex}
\title{Quantum magnetism with multicomponent 
dipolar molecules in an optical lattice}
\author{Ryan Barnett$^1$, Dmitry Petrov$^{1,2}$, Mikhail Lukin$^1$, and Eugene Demler$^1$}
\affiliation{$^1$Department of Physics,
  Harvard University, Cambridge, Massachusetts 02138, USA\\
$^2$ITAMP, Harvard-Smithsonian Center for Astrophysics, Cambridge, 
Massachusetts 02138, USA}
\date{\today}
\begin{abstract}
We consider bosonic dipolar molecules in an optical lattice
prepared in a mixture of different rotational states.
The $1/r^3$ interaction between molecules for this system is produced
by exchanging a quantum of angular momentum between two molecules.
We show that the Mott states of such
systems have a large variety of quantum phases characterized by dipolar
orderings including a 
state with ordering wave vector that can be changed by tilting the
lattice.  As the Mott insulating phase
is melted, we also describe several exotic superfluid phases that will occur. 
\end{abstract}

\maketitle

In ultracold physics, systems with long-range dipolar
interactions have recently attracted considerable attention  
both theoretically and
experimentally (for a recent review of ultracold dipolar molecules 
see \cite{Doyle04} and references therein). 
For atoms, dipolar interactions come from
their magnetic moments and become important for large
electronic spin \cite{Goral00}. Recent experiments demonstrated
the relevance of such dipolar interactions for the expansion of Cr atoms
from the BEC state \cite{Stuhler05}. On the other hand, for 
heteronuclear molecules, 
dipolar interactions arise
from their electric dipole moments.  Recent experiments have
succeeded in trapping and cooling several types of heteronuclear
molecules \cite{Doyle04, Stan04, Inouye04, Sage05}.
In a state with a well-defined angular momentum, molecules do
not have a dipole moment.  However, when an external electric field 
is used to polarize the molecules, dipolar moments can be induced. 
There has been considerable theoretical effort to study the resulting
dipole interactions and many-body physics associated with such systems
\cite{Yi00,Santos00,DeMille02,Goral02,Moore03,Micheli05}.

In this Letter, we consider an alternative
mechanism for obtaining the $1/r^3$ dipolar interactions, 
and the
important concomitant directional character.
Namely, we investigate a mixture of heteronuclear dipolar molecules in the
lowest (${\cal N}=0$) and the first excited (${\cal N}=1$) rotational states. 
For such a system, the origin of the long-range interaction is the exchange of
angular momentum quanta between molecules. 
We demonstrate that when
loaded  into an optical lattice, such mixtures can realize various
kinds of non-trivial effective dipolar spin systems with anisotropic, 
long-range
interactions. 
Several approaches for realizing spin systems using cold
atoms have been discussed before, including bosonic mixtures in
optical lattices in the Mott state
\cite{Mandel03,Duan03,Kuklov04, Isacsson05}, interacting fermions in
special lattices \cite{Damski05}, and trapped ions interacting with
lasers \cite{Deng05}.  The system we consider has the practical
advantages of the high energy scale for spin-dependent phenomena (set by
dipolar interactions) and the new physics associated with the
long-ranged nature of the dipole
interactions. Experimental realization of the system 
will give insight into several open questions in
condensed matter physics including competition between ferro and
antiferroelectric orders in crystals \cite{Sauer1940, Luttinger1946}
and systems with frustrated spin interactions \cite{Moessner03}.

Consider the system that contains
bosonic molecules in the lowest (${\cal N}=0, {\cal N}_z=0$) and first excited
(${\cal N}=1$,
${\cal N}_z=-1,0,1$) rotational states where we let $s^{\dagger}$ and
$t^{\dagger}_{-1,0,1}$ create these respective states.
We will often use the change of basis
$t^{\dagger}_{x}=(t^{\dagger}_{1}+t^{\dagger}_{-1})/\sqrt{2}$, 
$t^{\dagger}_{y}=-i (t^{\dagger}_{1}-t^{\dagger}_{-1})/\sqrt{2}$,
and $t^{\dagger}_z=t^{\dagger}_{0}$. To describe molecules in an
optical lattice we use the one-band Hubbard type effective model 
\begin{equation}
{\cal H} = {\cal H}_{\rm kin} + {\cal H}_{\rm Hub} +{\cal H}_{\rm
dip}.
\label{Eq:Hamiltonian}
\end{equation}
The first
term on the right hand side of (\ref{Eq:Hamiltonian}) is the
kinetic energy from nearest-neighbor hopping
${\cal H}_{\rm kin} = -J\sum_{\expect{ij}} 
\left(s^{\dagger}_{i}s_{j} +t_{i
\alpha}^{\dagger}t_{j \alpha}+{\rm h.c.}\right)$. 
Operators $s^{\dagger}_i$ and
$t_{i\alpha}^{\dagger}$ create molecules on
site $i$ (here and after the summation over repeating
indices $\alpha =x$, $y$, $z$ is implied). The last term in (\ref{Eq:Hamiltonian}) describes the dipolar
interaction between molecules from different sites
\begin{equation}
{\cal H}_{\rm dip}=\frac{\gamma}{2} \sum_{i \ne j}
\frac{d_{i \alpha}d_{j \alpha}
-3 d_{i \alpha} e_{ij\alpha} d_{j \beta} e_{ij\beta}}
{|{\bf R}_{i}-{\bf R}_{j}|^3}
\label{Eq:DipolarInt}
\end{equation}
where ${\bf R}_i$ are lattice vectors, $e_{ij\alpha}$
is the $\alpha$-component of the unit vector along 
${\bf R}_{i}-{\bf R}_{j}$, 
and parameter $\gamma$ equals $2d^2/3$, where $d$ is the
value of the dipole moment associated with the 
${\cal N}=0 \rightarrow {\cal N}=1$ transition, and
$d_{i}$ is the dipole moment operator at site $i$. 
The $\alpha$-component of the operator ${\bf d}_{i}$ is written
as
\begin{equation}
d_{i \alpha}=s_{i}^{\dagger} t_{i \alpha} e^{-2 i B_e t} + 
t_{i \alpha}^{\dagger} s_{i} e^{2 i B_e t},
\label{Eq:DipolarMoment}
\end{equation}
where we absorbed the energy difference between the rotational levels
$E_{{\cal N}=1}-E_{{\cal N}=0}=2B_e$ into the time dependence of the $t$ operators.
Since the rotational constant $B_e$ is considerably larger than any
other energy scale in the system, we assume that the terms in
(\ref{Eq:DipolarInt}) that oscillate at frequencies $\pm 4B_e$
average to zero. This forces the number of molecules in the ${\cal N}=0$ 
and ${\cal N}=1$ states to be independently conserved.  
Then
(\ref{Eq:DipolarInt}) reduces to
\begin{equation}
{\cal H}_{\rm dip}=\frac{\gamma}{2} \sum_{i \ne j}
\frac{(s^{\dagger}_{i} t_{j\alpha}^{\dagger} s_{j} t_{i\beta} + {\rm
h.c.})
(\delta_{\alpha \beta}-3  e_{ij\alpha}  e_{ij \beta})}
{|{\bf R}_{i}-{\bf R}_{j}|^3}.
\label{Eq:DipolarHamiltonian}
\end{equation}

The second term on the right hand side in (\ref{Eq:Hamiltonian})
is the Hubbard on-site interaction. For two $s$ molecules in the absence of an external electric field, the long-range part of their interaction potential is dominated by the van der Waals tail $C_6/R^6$ originating from second order terms in the
dipole-dipole interaction operator \cite{note}. For polar molecules with large
static rotational polarizabilities one can estimate $C_6\approx
-d^4/6B_e$. For the RbCs molecule ($d=0.5$ a.u., $B_e=7.7\times 10^{-8}$
a.u.) we have $C_6\approx 1.5\times 10^5$ a.u. For molecules with
smaller dipole moments and larger rotational constants like, for
example, CO ($d=0.043$, $B_e=9.0\times 10^{-6}$), the van der Waals
interaction is comparable in magnitude to interatomic forces. In any
case the range of the potential, which scales as $R_e=(mC_6)^{1/4}$,
is not much different from typical ranges of interatomic potentials
(for RbCs 
$R_e\approx 400$ a.u.). 
First order terms in the dipole-dipole operator are also absent for two molecules with ${\cal N}=1$. In this case, apart from a weak quadrupole-quadrupole contribution proportional to $R^{-5}$, the long-range part of the intermolecular potential is given by the van der Waals interaction with a comparable $C_6$ coefficient. Thus, the interactions between molecules with the same ${\cal N}$ are all short ranged and, in an ultracold system, can be modeled by contact potentials. Then, averaging them over the Gaussian on-site wave functions gives the Hubbard on-site interaction.   

The interaction between $s$ and $t_\alpha$ molecules (without loss of
generality we consider $\alpha=z$ here) is similar to the resonant
interaction of an electronically excited atom and a ground state
atom. For even partial waves the intermolecular potential
is asymptotically given by 
$W_z({\bf R})=\gamma(1-3\cos^2\theta_z)/2R^3$, 
where $\theta_z$ is the angle
between ${\bf R}$ and the $z$-axis. We consider the weakly
interacting regime where the characteristic energy scale of this
interaction, $\Delta E\sim\gamma l_0^{-3}$, is smaller than the Bloch
band separation.
Here $l_0$ is the oscillator length of the on-site harmonic
confinement. Then, the two-body problem in a harmonic potential can
be solved in the mean-field approximation by using the
pseudopotential approach (see \cite{You04} and references therein). Due to the anisotropy of $W_z({\bf R})$ the corresponding on-site interaction
energy, $V_z$, can be tuned at will by changing the aspect ratio of
the on-site confinement \cite{You04}. 

We arrive at the following expression for ${\cal H}_{\rm Hub}$:
\begin{eqnarray}
&&{\cal H}_{\rm Hub}=\sum_{i} \biggl[
\frac{U}{2}n_{s i} (n_{s i}-1) +
\frac{U_\alpha}{2}n_{t_\alpha i} ( n_{t_\alpha
i}-1)\nonumber\\
&&+\sum_{\alpha\ne\beta}U_{\alpha\beta}n_{t_\alpha i}n_{t_\beta
i}+V_\alpha n_{s i}n_{t_\alpha i}
\biggr].
\label{Eq:Onsite}
\end{eqnarray}
It is easy to see that (\ref{Eq:Onsite}) holds for arbitrary filling factors as long as the on-site density profiles remain Gaussian. However, for the purpose of this paper it is sufficient to consider on average one or two molecules per site.
%
The ten coupling constants in Eq.~(\ref{Eq:Onsite} are difficult to control
independently. However, in many cases not all of them are relevant. For a
mixture of $s$ and $t_z$ molecules the relevant coupling constants are
$U$, $U_z$, and $V_z$. These are tunable through trap aspect ratio and/or
Feshbach resonance. When considering this system we will take $U=U_z$
and $V\equiv V_z$. Another example is the Mott insulating state with one
molecule per site (all kinds of molecules are allowed) when the
interaction energies are much larger than $J$. Then, the particular values
of the on-site coupling constants are not important (for mechanical
stability it is sufficient that they are positive) and the state of the
system is found by minimizing the intersite dipolar interactions.
%


We now discuss the resulting phase diagram, first
focusing our attention on the Mott insulating state
with one molecule per site.
We take the variational wave function 
\begin{equation}
\ket{\Psi_{\rm MI}}=\prod_{i} \left(\cos(\theta)
s_{i}^{\dagger} + 
\sin(\theta)\psi_{i \alpha} t_{i \alpha}^{\dagger}\right) \ket{0}
\label{Eq:MottWavefunction}
\end{equation}
where $\theta$ describes the fraction of the molecules excited into
${\cal N}=1$ states and $\psi_{i \alpha}$ is a normalized complex vector
$\psi_{i \alpha}^* \psi_{i \alpha}$=1. 
Here, $\psi_{i \alpha}$ is
the variational parameter which descibes the direction the dipole moment
points on site $i$.
%
This allows
us to construct variational states that benefit maximally from dipolar
interactions. In all cases discussed below we verified the absence of
phase separation by checking the eigenvalues of the compressibility
matrix for $s$ and $t$ bosons
\cite{Huang87}.
Taking the expectation value of the dipole operator 
(\ref{Eq:DipolarMoment}) 
with our
variational wave function (\ref{Eq:MottWavefunction})
we obtain
$
\expect{d_{i \alpha}}=\sin(2\theta) |\psi_{\alpha}| \cos(\varphi_{i
\alpha}-2 B_e t)
$
where we have written $\psi_{i \alpha}=|\psi_{\alpha}|e^{i\varphi_{i
\alpha}}$.
Upon taking the expectation value of 
dipole Hamiltonian (\ref{Eq:DipolarHamiltonian}),
we find for the dipolar energy
\begin{eqnarray}
\label{Eq:SimplifiedEnergyDipolar}
E_{\rm dip} &=& \frac{\gamma \sin^2(2\theta)}{8} 
\\
\nonumber
&\times&\sum_{i \neq j}
\frac{(\psi_{i\alpha} \psi^{*}_{j \beta}+{\rm c.c.})
(\delta_{\alpha\beta} - 3 e_{ij\alpha} e_{ij\beta})}
{|{\bf R}_{i}-{\bf R}_{j}|^3}.
\end{eqnarray} 

When minimizing the energy in (\ref{Eq:SimplifiedEnergyDipolar})
it is important to keep track of the conservation laws that may be
present for certain experimental geometries and on the initial preparation
of the system. 
We will now consider several examples of ordering in the Mott
insulating state.
Although the dipole interaction in all cases is
described by (\ref{Eq:SimplifiedEnergyDipolar}) we will see that
different preparation leads to very different types of order.
Though all discussion in this work will be restricted to 2d,
we emphasize that there are nontrivial results in the Mott insulating
phase for the 1d and 3d cases as well.
As the first example, we consider the square lattice in 
the $xy$-plane defined by
vectors ${\bf a}_1={\bf \hat{x}}$ and  ${\bf a}_2={\bf \hat{y}}$.
Due to cross-terms such as 
$s^{\dagger} t^{\dagger}_{x} s t_{y}$
in the dipolar hamiltonian (\ref{Eq:DipolarHamiltonian}), we see that
$t_x$ molecules
can be converted to $t_y$ molecules and vice-versa.
Thus, $N_{t_x}$ and $N_{t_y}$ are not conserved quantities, and,
consequently, the only conserved quantities are 
$N_{s}$ and $N_{t_z}$.  Now consider preparing this system
in a  mixture of ${\cal N}=0$ and ${\cal N}=1, {\cal N}_z=1$ states.
Then after the system relaxes,
taking the constraints into account, we must have fixed
$\expect{N_{s}}=N \cos^2(\theta)$, 
$\expect{N_{t_x}}+\expect{N_{t_{y}}}=N \sin^2(\theta)$,
and $\expect{N_{t_z}}=0$.
This gives the constraints on the variational
wave function $\psi_{iz}=0$ and
$|\psi_{x}|^2+|\psi_{y}|^2=1$.
We see that the dipoles are allowed to rotate freely in
the $xy$-plane.  For this case, the dipoles will choose
to point head-to-tail in the direction of one of the bonds, 
while alternating in the other.  Thus, it is straightforward to see that
this gives the \emph{ordering wave vector} ${\bf q}=(0,\pi,0)$ with 
$\psi_{ix}=e^{i ({\bf q}\cdot {\bf R}_{i}+\varphi_0)}$ and 
$\psi_{iy}=\psi_{iz}=0$ where $\varphi_0$ is an arbitrary
phase corresponding to a change of phase of the time dependent oscillations
of the dipolar moment. 

We point out that this configuration is degenerate to the one
with dipoles pointing head-to-tail in the $y$-direction.

As the next example in two dimensions,
we take the same lattice as in the previous example, but prepare the
system in a mixture of ${\cal N}=0$ and ${\cal N}=1, {\cal N}_z=0$ states.  
Recalling that for this
geometry, both $N_s$ and $N_{t_z}$ are conserved quantities,
we find the constraint on the variational wave function
$\psi_{ix}=\psi_{iy}=0$ and 
$\psi_{iz}=e^{i \varphi_i}$.
With this constraint, the dipole interaction energy is
\begin{equation}
E_{\rm dip} = \frac{\gamma  \sin^2(2\theta)}{4} 
\sum_{i \neq j}
\frac{\cos(\varphi_{i}-\varphi_{j})}
{|{\bf R}_{i}-{\bf R}_{j}|^{3}}.
\end{equation}
Here, the dipoles are confined to point in the $z$-direction,
and therefore cannot point head-to-tail.  This gives antiferromagnetic
ordering in all directions, ${\bf q}=(\pi,\pi,0)$, with 
$\psi_{iz}=e^{i ({\bf q}\cdot {\bf R}_i+\varphi_0)}$ where
$\varphi_0$ is an arbitrary phase.

%
\begin{figure}[h]
\includegraphics[width=3.5in]{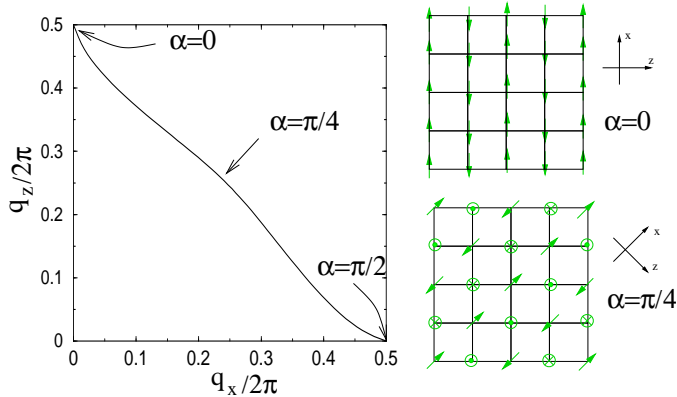}
\caption{The ordering wave vector as the lattice is
tilted by angle $\alpha$.  As described in the text,
for this situation a magnetic field is used to break the
degeneracy between (${\cal N}=1$, ${\cal N}_z=-1,0,1$) states.}
\label{Fig:spinsquare}
\end{figure}

For the final example for the Mott insulating state,
we consider a lattice in the $xz$-plane given by 
${\bf a}_1=\cos(\alpha){\bf \hat{x}}+\sin(\alpha){\bf \hat{z}}$
and
${\bf a}_2=-\sin(\alpha){\bf \hat{x}}+\cos(\alpha){\bf \hat{z}}$.
In addition,
we consider breaking the degeneracy of the 
(${\cal N}=1$, ${\cal N}_z=-1,0,1$) states with an external static magnetic
field in the $z$-direction which will introduce the term 
proportional to $B L_{z}$ into our hamiltonian.  
Preparing the
system in a superposition of ${\cal N}=0$ and ${\cal N}=1, {\cal N}_z=1$ states,
we note that because of this degeneracy breaking, there
will be no mixing between other angular momentum states.
That is, we can completely neglect the $t_{-1,0}$ states.
This will give $\psi_{ix}=-i\psi_{iy}=e^{\varphi_{i}}/\sqrt{2}$
and $\psi_{z}=0$ which 
will confine our dipoles to rotate in the $xy$-plane as:
$
\expect{{\bf d}_{i}(t)} = d_0 \cos (\varphi_i-2 B_e t){\bf \hat{x}}
+d_0 \sin (\varphi_i-2 B_e t){\bf \hat{y}}.
$
The dipolar energy of this system is therefore
\begin{equation}
E_{\rm dip} = \frac{\gamma \sin^2(\theta)}{8} 
\sum_{i \neq j}
\frac{\cos(\varphi_i-\varphi_j)(1-\frac{3}{2}e_{ijx}^2)}
{|{\bf R}_{i}-{\bf R}_{j}|^{3}}.
\end{equation}
We use the ansatz $\varphi_i={\bf q}\cdot {\bf R}_i +\varphi_0$ 
to find the minimum of this dipolar energy for a particular lattice defined
by the angle $\alpha$, and the results are summarized in
Fig.~\ref{Fig:spinsquare}.


We now consider melting the Mott insulator,
and entering the superfluid (SF) state.  
An interesting question to consider is what happens
to the ordering wave vector as the Mott insulating state is
melted?  For instance, deep in the superfluid phase, we will 
have $q=0$ which is favorable for Bose-Einstein
condensation while we saw that antiferromagnetic ordering is 
typically favored in the Mott insulating state by dipolar interactions.  
One possibility 
is that the wave vector interpolates
smoothly between these two extremes as the hopping $J$ increases.  
Another possibility is
that the molecules in the $s$ and $t$ states phase-separate.
We will show below that both scenarios are possible depending on
on-site energy parameters in our original hamiltonian.
For simplicity, we restrict
our attention to the third example we discussed above for 
the Mott insulating state which was
a two dimensional lattice in the $xy$
plane prepared with $\sigma_z$ polarized light.  For further
simplicity, we take 
$\expect{N_{s}}=\expect{N_{t_z}}=N/2$.  As we saw before,
we can neglect populating the $t_x$ and $t_y$ states, and
this phase has antiferromagnetic ${\bf q}=(\pi,\pi,0)$
order in the Mott insulating
phase.

Allowing for noninteger occupation per site
motivates the variational wave function
\begin{equation}
\label{Eq:SFWavefunction}
\ket{\Psi}=\prod_{i} 
\left(\sum_{n=0}^{\infty}
\alpha_{n}\frac{(a_{i}^{\dagger})^n}{\sqrt{n!}}\right)\ket{0}
\end{equation}
where 
$
a_{i}^{\dagger}=\cos(\theta)e^{i {\bf p}_s \cdot {\bf R}_i
}s_{i}^{\dagger}+
\sin(\theta)e^{i {\bf p}_t \cdot {\bf R}_i}t_{iz}^{\dagger}
$
and normalization requires $\sum_{n} |\alpha_n|^2=1$ (compare
with (\ref{Eq:MottWavefunction})).
As before, this wave function maximizes the dipole energy
for a given site which is energetically favorable.
We can now use a canonical transformation to write our original
hamiltonian in terms of the boson operators $a_{i}^{\dagger}$ 
(defined above)
and 
$
b_{i}^{\dagger}=-\sin(\theta)e^{i {\bf p}_s \cdot {\bf R}_i
}s_{i}^{\dagger}+
\cos(\theta)e^{i {\bf p}_t \cdot {\bf R}_i}t_{iz}^{\dagger}
$
(a new variable resulting from the 
transformation), and drop
the terms which give zero when evaluated using the above
variational wave function (\ref{Eq:SFWavefunction}).  
This leads to the following
single-site mean field hamiltonian
\begin{widetext}
\begin{eqnarray}
{\cal H}_{\rm MF}=
&-2J&\sum_{\alpha=x,y}\sqrt{\cos^{4}(\theta)+\sin^{4}(\theta)+
2\cos^{2}(\theta)\sin^{2}(\theta)\cos(q_\alpha)}
\left(a^{\dagger} \expect{a}+a
\expect{a^{\dagger}}-\expect{a^{\dagger}}\expect{a}\right)
\\
&+&\frac{\gamma}{4} \sin^{2}(2\theta)(2n_a \expect{n_a}- \expect{n_a}^2)
\sum_{{\bf R}_i \ne 0} \frac{\cos({\bf q}\cdot {\bf R}_i)}{|{\bf
R}_i|^3}
+ \frac{1}{2} U n_a(n_a-1) + \frac{1}{4} (V-U) \sin^{2}(2\theta)n_a(n_a-1)
\nonumber
\end{eqnarray}
\end{widetext}
where $n_{a}=a^{\dagger} a$ and we have already performed the minimization
over the center of mass momentum ${\bf p}=({\bf p}_t+{\bf p}_s)/2$.
The ground state of this hamiltonian
for fixed $\theta$ (relative concentrations) and 
${\bf q}={\bf p}_t-{\bf p}_s$ (relative momentum)
can be determined self-consistently in $\expect{a}$
and $\expect{n_a}$ through iteration numerically.  The general
approach will
then be to minimize these ground state energies over $q_{x,y} \in
[0,\pi]$
and $\theta \in [0,\pi/2]$.  When the minimum occurs for 
$\theta \ne \pi/4$, phase separation will occur. 
 
\begin{figure}
\includegraphics[width=3.2in]{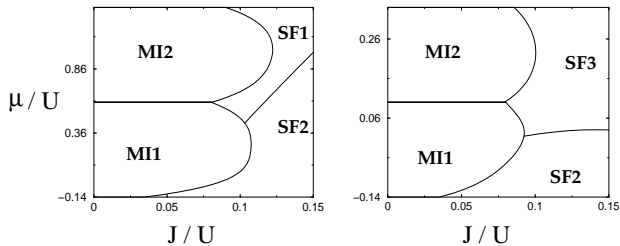}
\caption{The phase diagram for $V=U$ (left)
and $V=0$ (right).  For both cases, the dipolar
interaction strength was fixed at  
$\gamma=U/5$.  Shown are the antiferromagnetic MI states with
one and two bosons per site labeled MI1 and MI2.  
SF1 and SF2 correspond to superfluid states with partial
and complete phase separation (described in text). 
SF3 is a superfluid phase with no phase separation which has an ordering
wave vector that interpolates between the Mott insulating and deep superfluid
regime.}
\label{Fig:phased}
\end{figure}

The resulting phase diagrams are shown in Fig.~\ref{Fig:phased}.
The Mott insulating phases are antiferromagnetically aligned 
and were discussed in the previous section.  SF1 corresponds to
partial phase separation where part of the lattice will 
have a larger concentration of $s$ molecules while the other part
will have a higher concentration of $t$ molecules.  
Recall
that phase separation will occur for when $\theta \ne \pi/4$ since
we initially prepare the system to have equal populations of
molecules in the $s$ and $t_z$ states.  
The region
with more $s$ molecules will have $(p_s)_{x,y}=0$ and $(p_t)_{x,y}=\pi$.
This
will allow 
the more populated $s$ species to benefit maximally from BEC 
which prefers zero wave vector while still giving
$q_{x,y}=\pi$ which is preferred for the dipole interaction.
The similar situation holds for the region of the lattice with
a higher concentration of $t_z$ molecules.  SF2 corresponds to
the case where the $s$ and $t_z$ molecules completely phase separate.
Since the dipole interaction is negligible for this case,
we will have $(p_s)_{x,y}=(p_{t})_{x,y}=0$ which will favor BEC.  Finally,
SF3 corresponds to the case mentioned above where the wave vector
$\bf q$ interpolates between the deep superfluid and Mott insulating
states ($0<q_{x,y}<\pi$) for which no phase separation occurs 
($\theta=\pi/4$).

\comment{
We now briefly comment on a potential experimental detection 
schemes for such exotic superfluid and Mott insulating phases.  
The standard method of detection the so-called time-of-flight method which
requires a free expansion of the
molecular cloud.  This allows one to directly image the 
momentum density profile.  However, this cannot be immediately
performed for the dipolar gas since it is strongly interacting
with a $1/r^3$ potential which prohibits such a free expansion.  
One method out of this difficulty is to further excite
the molecules in the ${\cal N}=1$ rotational states into the ${\cal N}=2$
states, leaving a mixture of molecules in the ${\cal N}=0$ and ${\cal N}=2$
rotational states.  Since there will be no dipole interactions for
such a system, approximately free expansion will occur.
This will then allow one to measure the ordering wave vector
$\bf q$, for example in the antiferromagnetic state, by the
time-of-flight method.
}

In conclusion, we have shown that polar molecules prepared in a mixture of two
rotational states can exhibit long-range dipolar interactions in the absence
of an external electric field.  We have described several novel Mott insulating
and superfluid phases that can be realized as a result of such an interaction.
Such states can be detected by Bragg scattering or by time-of-flight
expansion \cite{Altman03}.

This work was supported by the NSF grant DMR-0132874 and career program, 
the Harvard-MIT CUA,
and the Harvard-Smithsonian ITAMP. 
We thank B. Halperin, R. Krems, A. Polkovnikov, D.W. Wang, and P. Zoller
for useful discussions.

\emph{Note added:}  When this manuscript
was close to completion we became aware of a paper considering a
similar system \cite{Micheli05}.


\begin{thebibliography}{10}
\bibitem{note}
\bibinfo{note}{
The rotational relaxation can be greatly reduced in an 
optical lattice,
similar to the inelastic losses in the atomic case. It can
completely eliminated in the 1 atom Mott case  and
suppressed in a deep lattice.}
\end{thebibliography}

\end{document}